\documentclass[fleqn,usenatbib]{mnras}
\usepackage{amsmath,amssymb,amsbsy}
\usepackage{graphicx}
\usepackage{epstopdf}
\usepackage{setspace}
\usepackage{flushend}
\usepackage{hyperref}
\usepackage[T1]{fontenc}
\usepackage{ae,aecompl}
\hypersetup{breaklinks=true} 

\title[Observational constraints on Tsallis cosmology]{Observational constraints on Tsallis modified gravity}
\author[Asghari and Sheykhi]{
    Mahnaz Asghari,$^{1}$\thanks{mahnaz.asghari@shirazu.ac.ir}
    Ahmad Sheykhi$^{1}$\thanks{asheykhi@shirazu.ac.ir}
    \\
    $^{1}$Physics Department, College of Sciences, Shiraz University, Shiraz 71454, Iran and \\
    Biruni Observatory, College of Sciences, Shiraz University, Shiraz 71454, Iran
    }

\date{Accepted XXX. Received YYY; in original form ZZZ}
\pubyear{2021}

\begin{document}
\label{firstpage}
\pagerange{\pageref{firstpage}--\pageref{lastpage}}
\maketitle

\begin{abstract}
The thermodynamics-gravity conjecture reveals that one can derive
the gravitational field equations by using the first law of
thermodynamics and vice versa. Considering the entropy associated
with the horizon in the form of non-extensive Tsallis entropy,
$S\sim A^{\beta}$ here we first derive the corresponding
gravitational field equations by applying the Clausius relation
$\delta Q=T \delta S$ to the horizon. We then construct the
Friedmann equations of Friedmann-Lema\^itre-Robertson-Walker
(FLRW) universe based on Tsallis modified gravity (TMG). Moreover,
in order to constrain the cosmological parameters of TMG model, we
use observational data, including Planck cosmic microwave
background (CMB), weak lensing, supernovae, baryon acoustic
oscillations (BAO), and redshift-space distortions (RSD) data.
Numerical results indicate that TMG model with a
quintessential dark energy is more compatible with
the low redshift measurements of large scale structures
by predicting a lower value for the structure growth parameter
$\sigma_8$ with respect to $\Lambda$CDM model.
This implies that TMG model would slightly alleviate the $\sigma_8$ tension.
\end{abstract}
\section{Introduction}
Considering the discovery of black hole thermodynamics in $1970 $
decade \citep{bh1,bh2,bh3}, it has been known that there is a
remarkable analogy between thermodynamics and gravity.
Accordingly, there should be some deep relationship between
thermodynamics and gravitational field equations, which first
disclosed by Jacobson in 1995 \citep{j1}. He derived the covariant
form of the Einstein field equations by using the Clausius
relation $\delta Q=T \delta S$, together with proportionality of
entropy to the horizon area. Jacobson's investigation reveals that
Einstein equations of general relativity is nothing but an
equation of state for the spacetime. Also there are more studies
in order to explore the profound connection between the theory of
general relativity and the laws of thermodynamics
\citep{grth1,grth2,grth3,grth4,grth5,grth6,grth7,grth8,grth9}.
Moreover, it is possible to rewrite the Friedmann equations in the
form of the first law of thermodynamics at the apparent horizon of
FLRW universe and vice versa
\citep{fr1,fr2,fr3,fr4,fr5,fr6,fr7,fr8}. The relationship
between gravity and thermodynamics is also considered in the
context of braneworld cosmology \citep{br1,br2,br3,br4}.

Although there are interesting proposals on disclosing the
relationship between gravity and thermodynamics, the nature of
this deep connection is not clearly explained. To this aim,
Verlinde proposed gravity as an entropic force (rather than a
fundamental one) caused by a change in the amount of information
associated with the positions of material bodies \citep{v1}.
Following this holographic scenario, he derived the Newton's law
of gravity and then extended results to relativistic case which
yield the Einstein equations. There are more investigations on
Verlinde's entropic force in the literatures
\citep{v2,v3,v4,v5,v6,v7,v8,v9,v10,v11,v12,v13,v14,v15,v16,v17}. On
the other hand, Padmanabhan considered spacetime as an emergent
structure (rather than a pre-existing background manifold) in
addition to treating field equations as the equations of emergent
phenomena \citep{p1}. He discussed the spatial expansion of the
universe as emergence of space as cosmic time progresses. One can
find related explorations on Padmanabhan's approach in Ref.
\citep{p2,p3,p4,p5,p6,p7,p8}.

In the cosmological setup, it has been shown that one can derive
the Friedmann equation of FLRW universe from the first law of
thermodynamics $\mathrm{d}E=T\mathrm{d}S+W\mathrm{d}V$ at apparent
horizon \citep{fr7}. In this relation, $S$ is the entropy
associated with the apparent horizon, which has the same
expression as the entropy of the black hole in the background
gravity theory. The only change needed is replacing the black hole
horizon radius $r_{+}$ with the apparent horizon radius
$\tilde{r}_\mathrm{A}$ in the entropy associated with the apparent
horizon. Accordingly, $T$ is the associated temperature with the
apparent horizon, $E$ is the total energy content of the universe
and $V$ is the volume inside the apparent horizon, and $W$ is the
work density defined as $W=\frac{1}{2}(\rho-p)$, where $\rho$ and
$p$ are the energy density and pressure of matter and energy
content in the universe, respectively.

The most familiar relation for the black hole entropy is the
Bekenstein-Hawking area law entropy \citep{bh4,bh3}
 \begin{equation} \label{eq1}
 S_{BH}=\frac{A}{4G} \;,
 \end{equation}
where $A=4 \pi r_{+}^2$ is the black hole horizon area. It should
be noted that area law of black hole entropy does not always hold.
For example, two well-known quantum corrections to the area law
are logarithmic (arises from the loop quantum gravity) and
power-law (appears in dealing with the entanglement of quantum
fields inside and outside the horizon) corrections which have been
extensively investigated in the literatures
\citep{qc1,qc2,qc3,qc4,qc5,qc6,qc7,qc8,qc9,qc10,qc11}. On the other
hand, it was argued that the Boltzmann-Gibbs (BG) theory is not
convincing in divergent partition function systems including large
scales gravitational systems \citep{bg1}. As a result, the BG
additive entropy should be generalized to non-additive
(non-extensive) entropy for such systems
\citep{na1,na2,na3,na4,na5,na6}. In this regards, and using the
statistical arguments, Tsallis and Cirto argued that the
microscopic mathematical expression of the thermodynamical entropy
of a black hole does not obey the area law and should be modified
as \citep{ts1,ts2}
 \begin{equation} \label{eq2}
 S=\gamma A^{\beta} \;,
\end{equation}
where $\gamma$ is a constant, $\beta$ is called the non-extensive
or Tsallis parameter, and $A$ is the black hole horizon area. It
should be noted that in the limit $\beta\to1$ the area law entropy
expression will be recovered. For more investigations related to
Tsallis entropy, refer to e.g. \citep{ts3,ts4,ts5,ts6,ts7,ts8,ts9,ts10}.

In the context of general relativity, one can derive
Einstein field equations by applying the area law entropy
(\ref{eq1}) in Clausius relation \citep{grth3}, which describes
the concordance $\Lambda$CDM model. However, in spite of the fact
that $\Lambda$CDM model is properly confirmed by current
observational data \citep{lcdm1,lcdm2}, there are some
discrepancies between the Planck cosmic microwave background (CMB)
data \citep{p18}, and low redshift observations, chiefly local
measurements of the structure growth parameter $\sigma_8$
\citep{s8}, and also local determinations of the Hubble constant
$H_0$ \citep{H01,H02,H03,H04}. These tensions might indicate some
new physics beyond the standard cosmological model. In this
direction, we investigate the potentiality of Tsallis modified
gravity (TMG) in alleviating the mismatch between local
observations and high redshift measurements.

In order to derive the modified Friedmann equations from Tsallis
entropy at apparent horizon of FLRW universe, we should consider
$A$ in relation (\ref{eq2}) as the area of apparent horizon
defined as
\begin{equation} \label{eq3}
 A=4\pi \tilde{r}^2_\mathrm{A} \;,
 \end{equation}
 where $\tilde{r}_\mathrm{A}$ is the apparent horizon radius \citep{ra1}
 \begin{equation} \label{eq4}
 \tilde{r}_\mathrm{A}=\Big(H^2+\frac{K}{a^2}\Big)^{-1/2} \;,
 \end{equation}
in which $H$ is the Hubble parameter and $K=-1,0,1$ is the
curvature constant corresponding to open, flat and closed
universe, respectively. Throughout this paper we set
$k_\mathrm{B}=c=\hbar=1$ for simplicity.

The paper is organized as follows. In section \ref{sec2}, we
derive the corresponding field equations as well as the modified
Friedmann equations in the context of TMG model. Section
\ref{sec3} is dedicated to numerical solutions of TMG model. In
section \ref{sec4} we constrain our model with observational data.
We summarize our results in section \ref{sec5}.
\section{Modified gravitational field equations from Tsallis entropy} \label{sec2}
In this approach, we consider a spatially flat 
homogeneous and isotropic FLRW universe, described
by the following line element in the background level
 \begin{equation} \label{eq5}
 \mathrm{d}s^2=a^2(\tau)\big(-\mathrm{d}\tau^2+\mathrm{d}\pmb{x}^2\big) \;,
 \end{equation}
 where $\tau$ is the conformal time.
Moreover, we are interested in scalar perturbations to linear
order, so following Ref. \citep{pt1}, the perturbed metric in
conformal Newtonian gauge is
    \begin{equation} \label{eq6}
    \mathrm{d}s^2=a^2(\tau)\Big(-\big(1+2\Psi\big)\mathrm{d}\tau^2+\big(1-2\Phi\big)\mathrm{d}\pmb{x}^2\Big) \;,
    \end{equation}
    where $\Psi$ and $\Phi$ are gravitational potentials. Also in synchronous gauge we have
    \begin{equation} \label{eq7}
    \mathrm{d}s^2=a^2(\tau)\Big(-\mathrm{d}\tau^2+\big(\delta_{ij}+h_{ij}\big)\mathrm{d}x^i\mathrm{d}x^j\Big) \;,
    \end{equation}
    in which $h_{ij}=\mathrm{diag}(-2\eta,-2\eta,h+4\eta)$, with scalar perturbations $h$ and $\eta$.
We assume the matter and energy content of the universe as a
perfect fluid with the following energy-momentum tensor
    \begin{equation} \label{eq8}
    T_{\mu \nu}=\big(\rho+p\big)u_{\mu}u_{\nu}+g_{\mu \nu}p \;,
    \end{equation}
where $\rho=\bar{\rho}+\delta\rho$ is the energy density,
$p=\bar{p}+\delta p$ is the pressure, and $u_{\mu}$ is the
four-velocity (and a bar indicates the background level).

In order to derive Einstein field equations from Tsallis entropy,
we apply the Clausius relation
    \begin{equation} \label{eq9}
    \delta Q=T \delta S \;.
    \end{equation}
It is assumed that the Clausius relation is satisfied on a local
causal horizon $\mathcal{H}$ (which here it is the apparent
horizon).

Let us now consider equation (\ref{eq9}) more specifically. The
associated temperature with the apparent horizon is given by
    \begin{equation} \label{eq10}
    T=\frac{\kappa}{2\pi} \;,
    \end{equation}
where $\kappa$ is the surface gravity at apparent horizon. Also,
according to equation (\ref{eq2}), $\delta S$ can be written as
    \begin{equation} \label{eq11}
    \delta S=\gamma \beta A^{\beta-1} \delta A \;.
    \end{equation}
In order to define $\delta Q$, we follow the approach of Ref.
\citep{j1,grth3} which reads
    \begin{equation} \label{eq12}
    \delta Q= -\kappa \int_{\mathcal{H}}^{} \lambda T_{\mu \nu} k^{\mu} k^{\nu} \mathrm{d}\lambda \mathrm{d}A \;,
    \end{equation}
and also we have
    \begin{align}
    & \delta A=\int_{\mathcal{H}}^{} \theta \mathrm{d}\lambda \mathrm{d}A \;, \label{eq13} \\
    & \theta=-\lambda R_{\mu \nu} k^{\mu} k^{\nu} \;. \label{eq14}
    \end{align}
Substituting relations (\ref{eq10}),(\ref{eq11}) and (\ref{eq12})
in equation (\ref{eq9}) would result in
    \begin{align}
    & \kappa \int_{\mathcal{H}}^{} (-\lambda) T_{\mu \nu} k^{\mu} k^{\nu} \mathrm{d}\lambda \mathrm{d}A \nonumber \\
    &=\frac{\kappa}{2\pi} \gamma \beta \int_{\mathcal{H}}^{} (-\lambda) R_{\mu \nu} k^{\mu} k^{\nu} A^{\beta-1} \mathrm{d}\lambda \mathrm{d}A \;, \label{eq15} \\
    & \to \int_{\mathcal{H}}^{} (-\lambda) \bigg(-2\pi T_{\mu \nu}+\gamma \beta R_{\mu \nu} A^{\beta-1} \bigg) k^{\mu} k^{\nu} \mathrm{d}\lambda \mathrm{d}A=0 \;. \label{eq16}
    \end{align}
So for all null $k^{\mu}$ we should have
    \begin{equation} \label{eq17}
    -2\pi T_{\mu \nu}+\gamma \beta R_{\mu \nu} A^{\beta-1}=f g_{\mu \nu} \;,
    \end{equation}
in which $f$ is a scalar. Then, the energy-momentum conservation
($\nabla^{\mu} T_{\mu \nu}=0$) would impose
    \begin{equation} \label{eq18}
    \nabla^{\mu} \big(\gamma \beta R_{\mu \nu} A^{\beta-1}-f g_{\mu \nu}\big)=0 \;,
    \end{equation}
which after doing some calculations results in
    \begin{equation} \label{eq19}
    \frac{1}{2} \gamma \beta \big(\partial_{\nu} R\big) A^{\beta-1} + \gamma \beta R_{\mu \nu} \partial^{\mu} A^{\beta-1} =\partial_{\nu} f \,.
    \end{equation}
However, the LHS of equation (\ref{eq19}) is not the gradient of a
scalar. It reveals that the Clausius relation does not hold due to
non-equilibrium thermodynamics. Hence, in order to resolve this
contradiction, we use the entropy balance relation \citep{grth3}
    \begin{equation} \label{eq20}
    \delta S=\frac{\delta Q}{T}+\mathrm{d}_i S \;,
    \end{equation}
where $\mathrm{d}_i S$ is the entropy produced inside the system
due to irreversible processes \citep{en1}. In order to restore the
energy-momentum conservation, we choose $\mathrm{d}_i S$ as
    \begin{equation} \label{eq21}
    \mathrm{d}_i S=\gamma \beta \int_{\mathcal{H}}^{} (-\lambda) \nabla_{\mu} \nabla_{\nu} A^{\beta-1} k^{\mu} k^{\nu} \mathrm{d}\lambda \mathrm{d}A \;.
    \end{equation}
Thus, equation (\ref{eq20}) reads
    \begin{align} \label{eq22}
    & \gamma \beta \int_{\mathcal{H}}^{} (-\lambda) R_{\mu \nu} k^{\mu} k^{\nu} A^{\beta-1} \mathrm{d}\lambda \mathrm{d}A \nonumber\\
    &=2\pi \int_{\mathcal{H}}^{} (-\lambda) T_{\mu \nu} k^{\mu} k^{\nu} \mathrm{d}\lambda \mathrm{d}A \nonumber\\
    &+\gamma \beta \int_{\mathcal{H}}^{} (-\lambda) \nabla_{\mu} \nabla_{\nu} A^{\beta-1} k^{\mu} k^{\nu} \mathrm{d}\lambda \mathrm{d}A \,,
    \end{align}
    \begin{align} \label{eq23}
    \to \int_{\mathcal{H}}^{} &(-\lambda) \bigg(\gamma \beta R_{\mu \nu} A^{\beta-1}-2\pi T_{\mu \nu}-\gamma \beta \nabla_{\mu} \nabla_{\nu} A^{\beta-1}\bigg) \nonumber\\
    & \times k^{\mu} k^{\nu} \mathrm{d}\lambda \mathrm{d}A=0 \;.
    \end{align}
Likewise, for all null $k^{\mu}$ we obtain
    \begin{equation} \label{eq24}
    \gamma \beta R_{\mu \nu} A^{\beta-1}-2\pi T_{\mu \nu}-\gamma \beta \nabla_{\mu} \nabla_{\nu} A^{\beta-1}=f g_{\mu \nu} \;.
    \end{equation}
According to the energy-momentum conservation, we should have
    \begin{equation} \label{eq25}
    \nabla^{\mu} \big(\gamma \beta R_{\mu \nu} A^{\beta-1}-\gamma \beta \nabla_{\mu} \nabla_{\nu} A^{\beta-1}-f g_{\mu \nu}\big)=0 \,.
    \end{equation}
Then, considering $\nabla^{\mu} \nabla_{\mu} \nabla_{\nu}
A^{\beta-1}=\nabla_{\nu} \Box A^{\beta-1}+R_{\zeta \nu}
\partial^{\zeta} A^{\beta-1}$, we get
    \begin{equation}  \label{eq26}
    \gamma \beta \Big(\frac{1}{2} \big(\partial_{\nu} R\big) A^{\beta-1}-\partial_{\nu} \Box A^{\beta-1}\Big)=\partial_{\nu} f \;.
    \end{equation}
It is possible to define the scalar $\mathcal{L}$ as
$\mathcal{L}=R A^{\beta-1}$, which results in
    \begin{align} \label{eq27}
    & \frac{\partial \mathcal{L}}{\partial R}=A^{\beta-1} \;, \nonumber\\
    & \partial_{\nu} R=\frac{\partial R}{\partial x^{\nu}}=\frac{\partial R}{\partial \mathcal{L}}\frac{\partial \mathcal{L}}{\partial x^{\nu}}=A^{1-\beta} \partial_{\nu}\mathcal{L} \;,  \nonumber\\
    & \to \big(\partial_{\nu} R\big) A^{\beta-1}=\partial_{\nu}\mathcal{L} \;.
    \end{align}
Thus equation (\ref{eq26}) takes the form
    \begin{equation}  \label{eq28}
    \gamma \beta \Big(\frac{1}{2}\partial_{\nu} \mathcal{L}-\partial_{\nu} \Box A^{\beta-1}\Big)=\partial_{\nu} f \;,
    \end{equation}
which results in the following equation for the scalar $f$
    \begin{align}  \label{eq29}
    f&=\gamma \beta\Big(\frac{1}{2} \mathcal{L}- \Box A^{\beta-1}\Big) \nonumber\\
    &=\gamma \beta\Big(\frac{1}{2} R A^{\beta-1} - \Box A^{\beta-1}\Big) \;.
    \end{align}
Therefore, the gravitational field equations based on nonextensive
Tsallis entropy (\ref{eq2}) take the following form
    \begin{align} \label{eq30}
    & R_{\mu \nu} A^{\beta-1}-\nabla_{\mu} \nabla_{\nu} A^{\beta-1}-\frac{1}{2} R A^{\beta-1} g_{\mu \nu}+\Box A^{\beta-1} g_{\mu \nu} \nonumber\\
    &=\frac{2\pi}{\gamma \beta}T_{\mu \nu} \;.
    \end{align}
Equation (\ref{eq30}) is indeed the modified Einstein field
equations, when the entropy associated with the horizon does not
obey the area law and instead is given by equation (\ref{eq2}). When
$\beta=1$, equation (\ref{eq30}) restores the standard Einstein
equations. Given the modified Einstein equations at hand, we are
in the position to derive the corresponding modified Friedmann
equations in the background of a flat FLRW universe. According to
relations (\ref{eq3}) and (\ref{eq4}), for a spatially flat
universe we have
    \begin{equation} \label{eq31}
    A^{\beta-1}=(4\pi)^{\beta-1} \Big(\frac{a'}{a^2}\Big)^{2-2\beta}=(4\pi)^{\beta-1} H^{2-2\beta} \;,
    \end{equation}
where a prime indicates a deviation with respect to the conformal
time, and $H={a'}/{a^2}$ is the Hubble parameter. The
$(00)$ and $(ii)$ components of the modified
field equations (\ref{eq30}) in a background level are given by
    \begin{align} \label{eq32}
    &\Big(\frac{a'}{a^2}\Big)^{2-2\beta}\Bigg{\{}\Big(\frac{a'}{a}\Big)^2+2\big(1-\beta\big)\bigg(\frac{a''}{a}-2\Big(\frac{a'}{a}\Big)^2\bigg)\Bigg{\}} \nonumber\\
    &=\frac{8\pi G}{3} a^2 \sum_{i}\bar{\rho}_i \;,
    \end{align}
    \begin{align} \label{eq33}
    &\Big(\frac{a'}{a^2}\Big)^{2-2\beta}\Bigg{\{}-2\frac{a''}{a}+\Big(\frac{a'}{a}\Big)^2+2\big(1-\beta\big)\Bigg[5\frac{a''}{a}
    -4\Big(\frac{a'}{a}\Big)^2 \nonumber\\
    &-\frac{a'''}{a'}-\big(1-2\beta\big)\bigg(\frac{a''}{a'}-2\frac{a'}{a}\bigg)^2\Bigg]\Bigg{\}}=8\pi G a^2 \sum_{i}\bar{p}_i \;,
    \end{align}
where we have taken the energy momentum tensor $T_{\mu\nu}$ as
defined in equation (\ref{eq8}), and also we have defined
$\gamma={(4\pi)^{1-\beta}}/{(4G\beta)}$. Here $i$ indicates the
component $i^{th}$ of the energy in the universe. In terms of the
Hubble parameter $H$, the above equations can be written as
    \begin{equation} \label{eq34}
    H^{4-2\beta}+2\big(1-\beta\big) H^{2-2\beta} H' \frac{1}{a}=\frac{8\pi G}{3} \sum_{i}\bar{\rho}_i \;,
    \end{equation}
    \begin{align} \label{eq35}
    & H^{2-2\beta}\Bigg{\{}2\big(\beta-2\big) H' \frac{1}{a}-3H^2 \nonumber\\
    &-2\big(1-\beta\big)\bigg(\big(1-2\beta\big)\frac{H'^2}{H^2}+\frac{H''}{H}\bigg)\frac{1}{a^2}\Bigg{\}}=8\pi G \sum_{i}\bar{p}_i \;.
    \end{align}
Combining equations (\ref{eq34}) and (\ref{eq35}), one can rewrite the
first modified Friedmann equation as
    \begin{equation} \label{eqF}
    H^{4-2\beta}=\frac{1}{4\beta-3}\frac{8\pi G}{3}\sum_{i}\bar{\rho}_i \;.
    \end{equation}
Taking into account the definition of the total density parameter
$\Omega_\mathrm{tot}={\bar{\rho}}/{\rho_\mathrm{cr}}$, where
$\rho_\mathrm{cr}={3H^2}/{(8\pi G)}$ and
$\bar{\rho}=\sum_{i}\bar{\rho}_i$, we can further rewrite equation
(\ref{eqF}) in the form
    \begin{equation} \label{eq36}
    \Omega_\mathrm{tot}=\big(4\beta-3\big)H^{2-2\beta} \;,
    \end{equation}
which differs from the standard cosmology, unless for the case
$\beta=1$ where we have $\Omega_\mathrm{tot}=1$.

It should be noted that in this model $\rho_\mathrm{tot}$ includes
radiation (R), matter (M) (dark matter (DM) and baryons (B)) and
the fluid "f" with a constant equation of state
$w_\mathrm{f}={\bar{p}_\mathrm{f}}/{\bar{\rho}_\mathrm{f}}$.
Considering  the total equation of state
$w_\mathrm{tot}={\bar{p}_\mathrm{tot}}/{\bar{\rho}_\mathrm{tot}}$,
under the condition \mbox{$w_\mathrm{tot}<({1-2\beta})/{3}$}, the
universe would experience an accelerated expansion. Choosing
$\beta=1$, results in $w_\mathrm{tot}<-{1}/{3}$ which is the
condition for an accelerated universe in standard cosmology.
However, different values of Tsallis parameter $\beta$, can result
in different conditions from general relativity. In
particular, considering $\beta\geq1/2$ would result in
$w_\mathrm{tot}<0$, while for $\beta<1/2$, $w_\mathrm{tot}$ can
take positive value and we still have the late-time acceleration
of the cosmic expansion. For example, by choosing $\beta=1/3$ we
obtain $w_\mathrm{tot}<1/9$. This implies that one may consider a
universe filled with baryonic matter, and still enjoys an
accelerated expansion without invoking any dark companion for its
matter/energy content. This is consistent with the argument given
in \citep{ts11}.

Moreover, modified field equations to linear order of
perturbations in conformal Newtonian gauge (con) take the form
    \begin{align} \label{eq37}
    &\Big(\frac{a'}{a^2}\Big)^{2-2\beta}\Bigg{\{}3\frac{a'}{a}\Phi'+k^2\Phi+3\Big(\frac{a'}{a}\Big)^2\Psi \nonumber \\
    &+
    3\big(1-\beta\big)\bigg(\frac{a''}{a'}-2\frac{a'}{a}\bigg)\bigg(\Phi'+2\Psi\frac{a'}{a}\bigg)\Bigg{\}} \nonumber \\
    &=-4\pi G a^2 \sum_{i}\delta \rho_{i(\mathrm{con})} \;,
    \end{align}
    \begin{align} \label{eq38}
    &\Big(\frac{a'}{a^2}\Big)^{2-2\beta}\Bigg{\{}k^2\Phi'+\frac{a'}{a}k^2\Psi+\big(1-\beta\big)
    \bigg(\frac{a''}{a'}-2\frac{a'}{a}\bigg)k^2\Psi\Bigg{\}} \nonumber \\
    &=4\pi G a^2 \sum_{i}\big(\bar{\rho}_i+\bar{p}_i\big)\theta_{i(\mathrm{con})} \;,
    \end{align}
    \begin{align} \label{eq39}
    \Phi=\Psi \;,
    \end{align}
    \begin{align} \label{eq40}
    &\Big(\frac{a'}{a^2}\Big)^{2-2\beta}\Bigg{\{}2\frac{a''}{a}\Psi-\Big(\frac{a'}{a}\Big)^2\Psi+\frac{a'}{a}\Psi'+2\frac{a'}{a}\Phi'
    +\Phi'' \nonumber \\
    &+\frac{k^2}{3}\big(\Phi-\Psi\big) +\big(1-\beta\big)\Bigg{\{}2\bigg(\frac{a''}{a'}-2\frac{a'}{a}\bigg)\bigg(2\frac{a'}{a}\Psi+\Phi'\bigg) \nonumber \\
    &+2\Psi\Bigg[\big(1-2\beta\big)\bigg(\frac{a''}{a'}-2\frac{a'}{a}\bigg)^2 +\frac{a'''}{a'}-7\frac{a''}{a}+8\Big(\frac{a'}{a}\Big)^2\Bigg] \nonumber \\ &+\Psi'\bigg(\frac{a''}{a'}-2\frac{a'}{a}\bigg)\Bigg{\}}\Bigg{\}}=4\pi G a^2 \sum_{i}\delta p_{i(\mathrm{con})} \;.
    \end{align}
Also in synchronous gauge (syn) we have
    \begin{align} \label{eq41}
    &\Big(\frac{a'}{a^2}\Big)^{2-2\beta}\Bigg{\{}\frac{a'}{a}h'-2k^2\eta+\big(1-\beta\big)h'\bigg(\frac{a''}{a'}-2\frac{a'}{a}\bigg)\Bigg{\}} \nonumber \\
    &=8\pi G a^2 \sum_{i}\delta \rho_{i(\mathrm{syn})} \;,
    \end{align}
    \begin{align} \label{eq42}
    \Big(\frac{a'}{a^2}\Big)^{2-2\beta} k^2\eta'=4\pi G a^2 \sum_{i}\big(\bar{\rho}_i+\bar{p}_i\big)\theta_{i(\mathrm{syn})} \;,
    \end{align}
    \begin{align} \label{eq43}
    &\Big(\frac{a'}{a^2}\Big)^{2-2\beta}\Bigg{\{}\frac{1}{2}h''+3\eta''+\frac{a'}{a}h'+6\frac{a'}{a}\eta'-k^2\eta \nonumber \\
    &+2\big(1-\beta\big)\bigg(\frac{a''}{a'}-2\frac{a'}{a}\bigg)\bigg(\frac{1}{2}h'+3\eta'\bigg)\Bigg{\}}=0 \;,
    \end{align}
    \begin{align} \label{eq44}
    &\Big(\frac{a'}{a^2}\Big)^{2-2\beta}\Bigg{\{}-2\frac{a'}{a}h'-h''+2k^2\eta \nonumber \\
    &-2\big(1-\beta\big)\bigg(\frac{a''}{a'}
    -2\frac{a'}{a}\bigg)h'\Bigg{\}}
    =24\pi G a^2 \sum_{i}\delta p_{i(\mathrm{syn})} \;.
    \end{align}
It can be easily seen that for $\beta=1$, the standard field
equations in Einstein gravity will be restored. In the following
we study TMG model in synchronous gauge.

Furthermore, the conservation equations in TMG model are the same
as those in general relativity, so for the matter and the fluid
components we find the following conservation equations in the
synchronous gauge
    \begin{align}
    \delta'_\mathrm{M(syn)}=-\theta_\mathrm{M(syn)}-\frac{1}{2}h' \;,
    \end{align}
    \begin{align}
    \theta'_\mathrm{M(syn)}=-\frac{a'}{a}\theta_\mathrm{M(syn)} \;,
    \end{align}
    \begin{align}
    &\delta'_\mathrm{f(syn)}=-3\frac{a'}{a}\big(c^2_{s,\mathrm{f}}-w_\mathrm{f}\big)\delta_\mathrm{f(syn)}-\frac{1}{2}h'\big(1+w_\mathrm{f}\big) \nonumber \\
    &-\big(1+w_\mathrm{f}\big)\bigg(1+9\Big(\frac{a'}{a}\Big)^2\big(c^2_{s,\mathrm{f}}-c^2_{a,\mathrm{f}}\big)\frac{1}{k^2}\bigg)\theta_\mathrm{f(syn)} \;,
    \end{align}
    \begin{align}
    \theta'_\mathrm{f(syn)}=\frac{a'}{a}\big(3c^2_{s,\mathrm{f}}-1\big)\theta_\mathrm{f(syn)}+\frac{k^2c^2_{s,\mathrm{f}}}{1+w_\mathrm{f}}\delta_\mathrm{f(syn)} \;.
    \end{align}

In order to understand the properties of TMG model more precisely
and also constrain the Tsallis parameter $\beta$ with
observational data, we modify the Boltzmann code Cosmic Linear
Anisotropy Solving System (CLASS) \citep{cl1} according to this
model. Furthermore, running a Markov Chain Monte Carlo (MCMC)
analysis through the M\textsc{onte} P\textsc{ython} code
\citep{mp1,mp2}, provides us with constraints on the cosmological
parameters.
\section{Numerical results} \label{sec3}
In this part, we study TMG model by implementing modified field
equations in the CLASS code. Accordingly, we use Planck 2018 data
\citep{p18} for the cosmological parameters given by
$\Omega_{\mathrm{B},0}h^2=0.02242$,
$\Omega_{\mathrm{DM},0}h^2=0.11933$,
$H_0=67.66\,\mathrm{km\,s^{-1}\,Mpc^{-1}}$, $A_s=2.105\times
10^{-9}$, and $\tau_\mathrm{reio}=0.0561$. Since the nature of
fluid is unknown to us, without loss of generality we can consider
$w_\mathrm{f}=-0.98$ and $c_{s,\mathrm{f}}^2=1$.

\begin{figure*}
        \includegraphics[width=8.5cm]{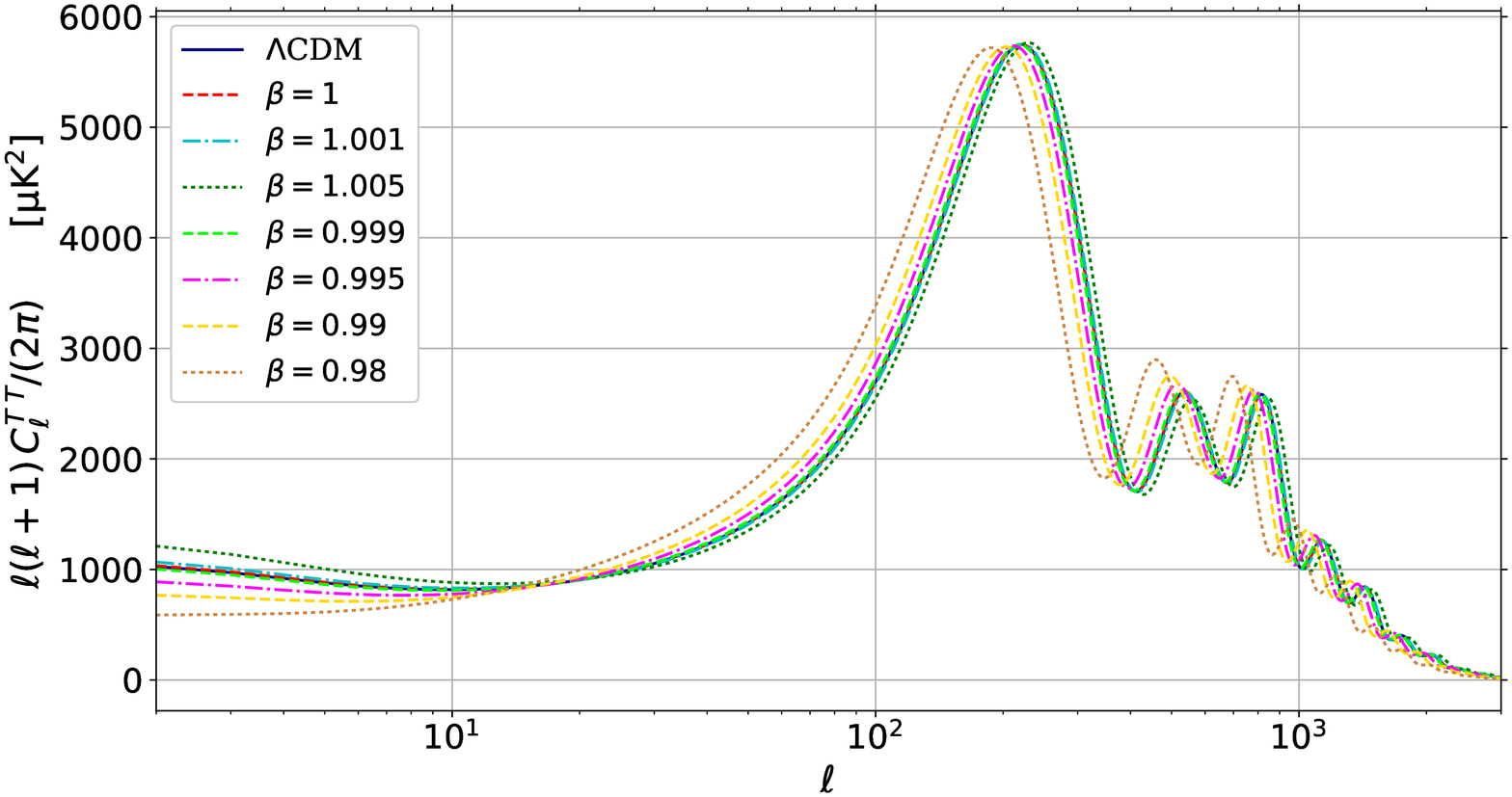}
        \includegraphics[width=8.5cm]{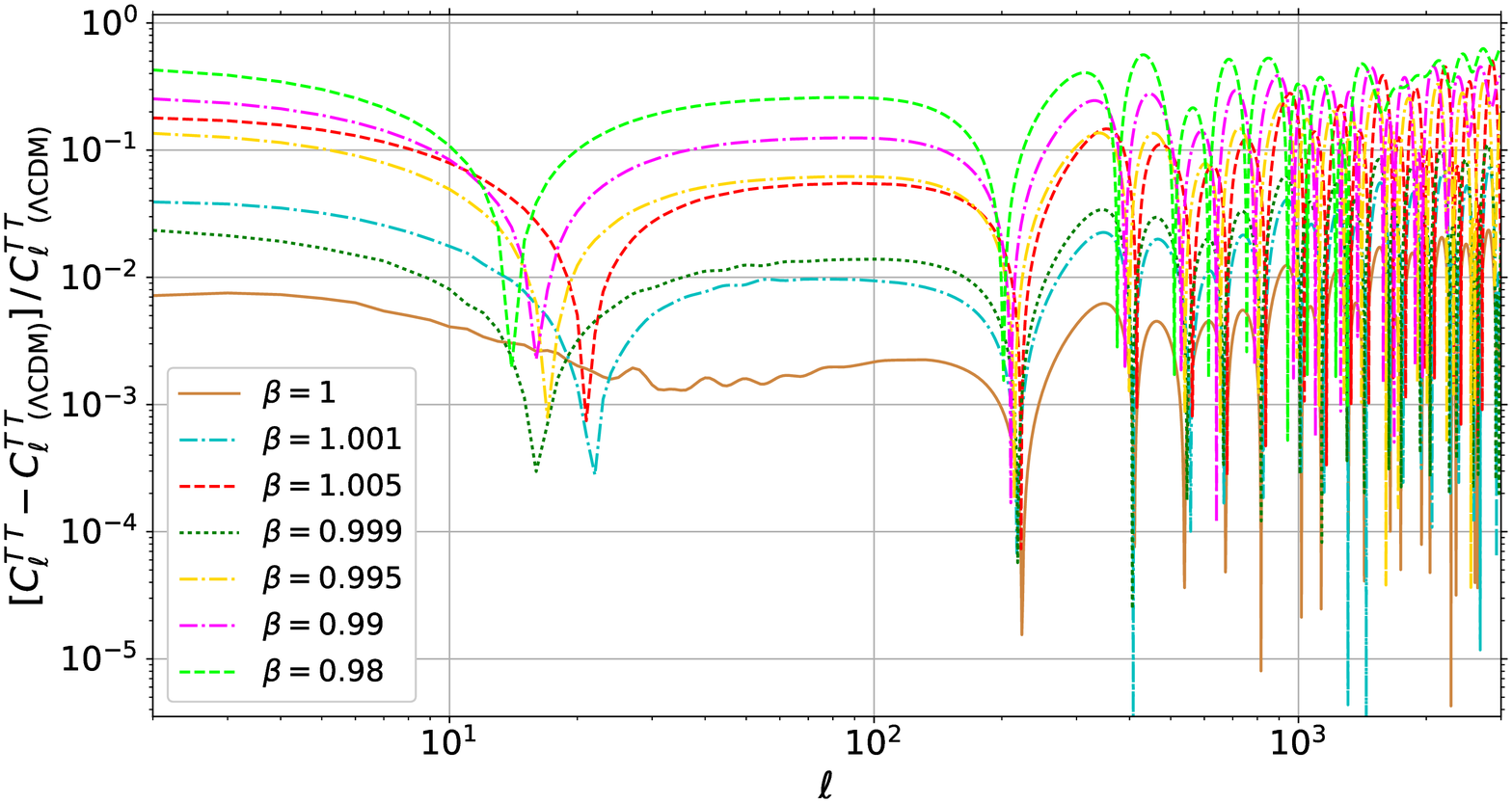}
        \includegraphics[width=8.5cm]{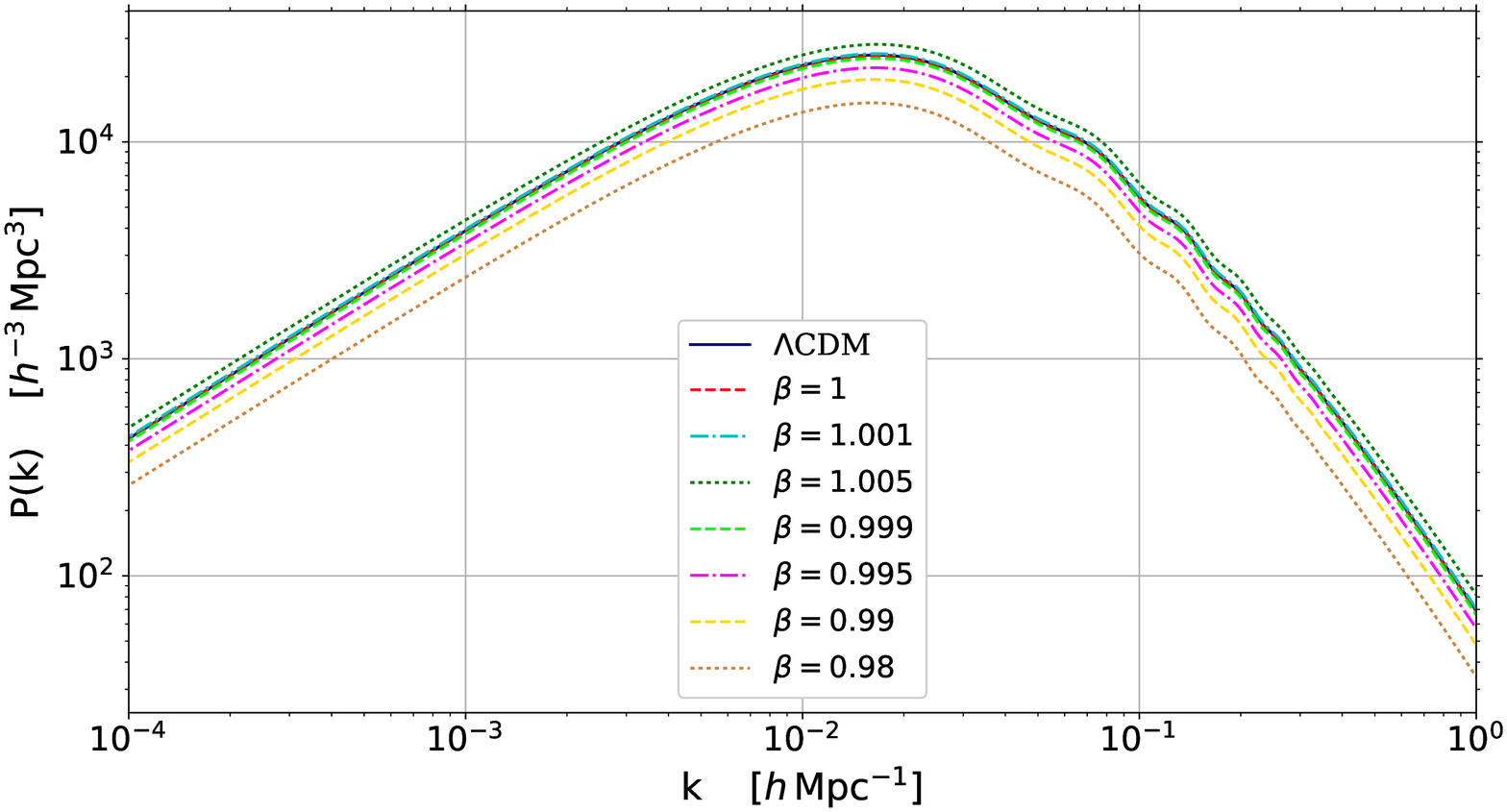}
        \includegraphics[width=8.5cm]{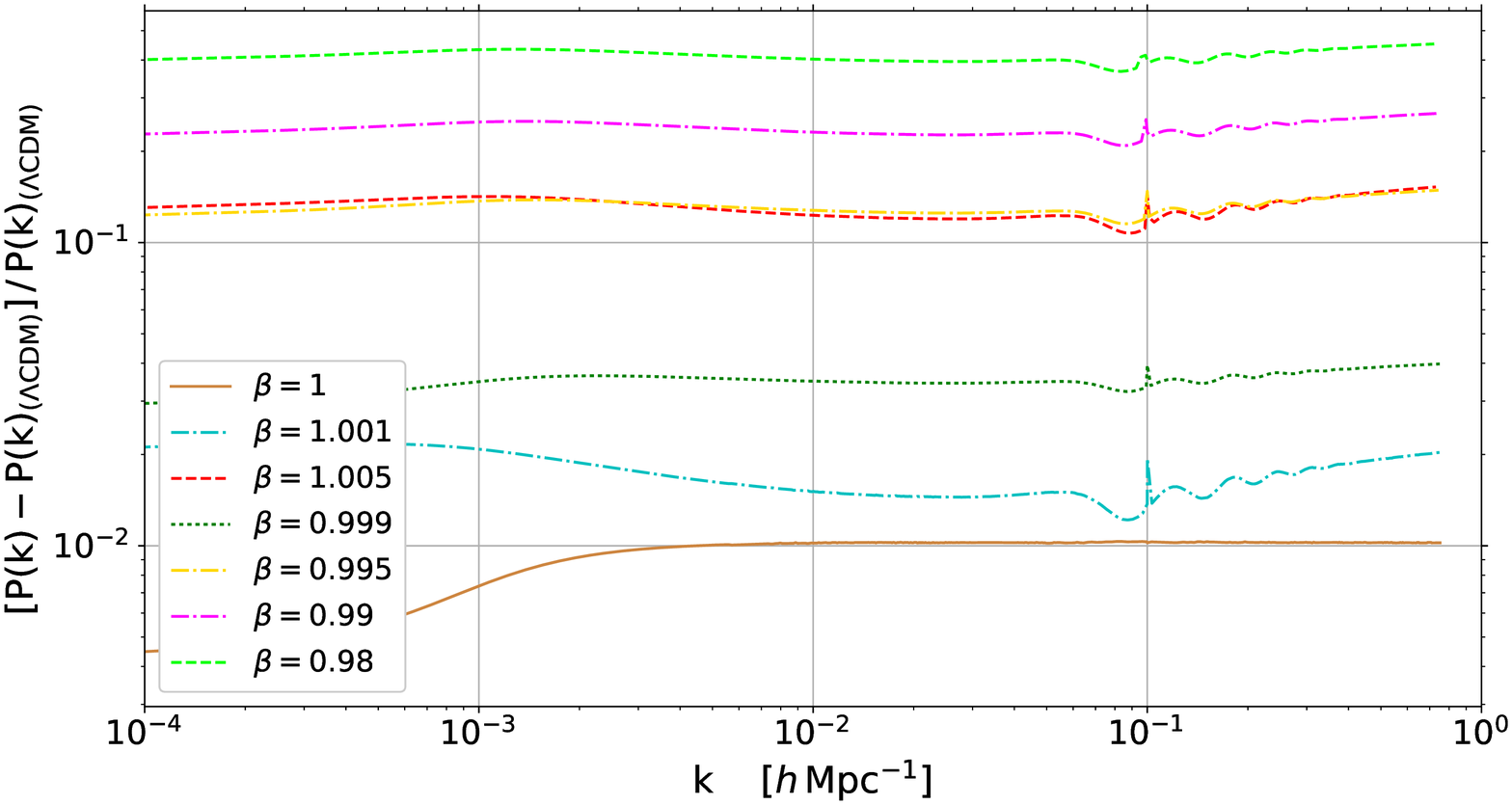}
        \caption{\small The CMB power spectra (upper left) and their relative ratio with
        respect to the standard model (upper right) for different values of $\beta$.
        Lower panels show the analogous diagrams for the matter power spectrum.}
        \label{f1}
\end{figure*}

\begin{figure}
    \centering
    \includegraphics[width=8.5cm]{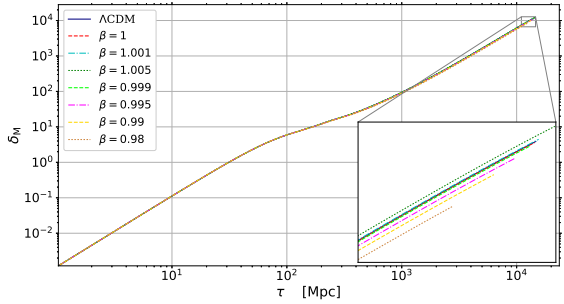}
    \caption{\small Matter density contrast diagrams in term of conformal time for TMG model with different values of $\beta$ compared to $\Lambda$CDM.}
    \label{f2}
\end{figure}

The CMB temperature anisotropy and
matter power spectra diagrams are shown in Fig. \ref{f1}.
According to matter power spectra diagrams, there is a decrease in
structure growth for TMG model with $\beta<1$.
So regarding discrepancies between low redshift observational data
and CMB results, it can be concluded that TMG model with $\beta<1$ is
more consistent with local measurements of the structure
growth parameter $\sigma_8$. This feature of TMG model
can be also understood from matter density contrast diagrams in
Fig. \ref{f2}.

\begin{figure}
    \centering
    \includegraphics[width=8.5cm]{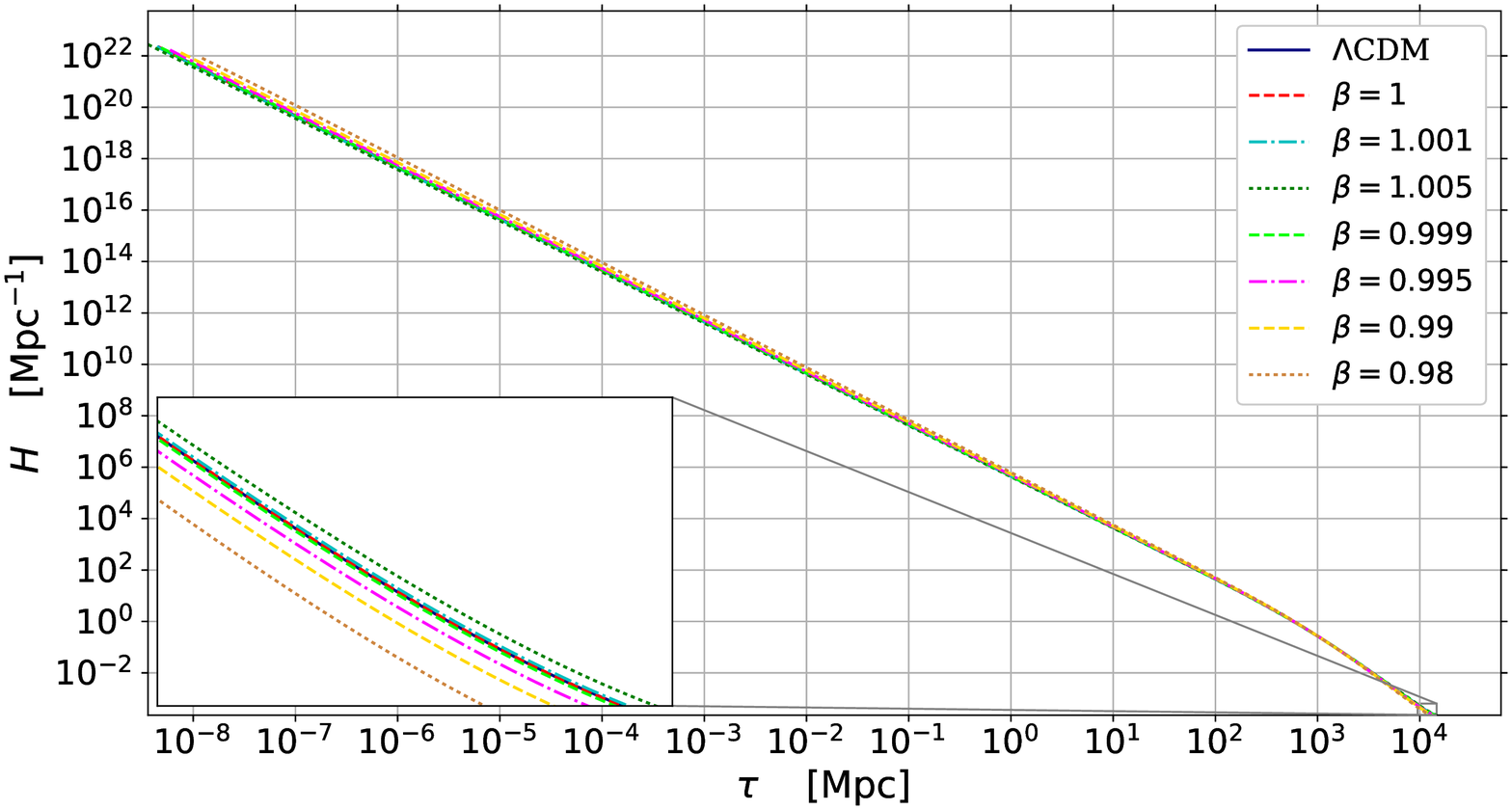}
    \caption{\small Hubble parameter in term of conformal time for TMG model with different values of $\beta$ compared to $\Lambda$CDM.}
    \label{fH}
\end{figure}

It is also interesting to investigate the evolution of the
Hubble parameter in TMG model which is described in equation
(\ref{eqF}). Considering the evolution of Hubble parameter
illustrated in Fig. \ref{fH}, it can be understood that choosing
$\beta>1$ would increase the current value of Hubble parameter,
which is consequently more compatible with local determinations of
this parameter. However, the $H_0$ tension problem becomes more
severe for $\beta<1$.
\section{Observational constraints} \label{sec4}
In order to explore constraints on cosmological parameters of TMG model,
we run an MCMC analysis using the M\textsc{onte} P\textsc{ython} code \citep{mp1,mp2}.
The following set of parameters is considered in the MCMC analysis: \\
\{$100\,\Omega_{\mathrm{B},0} h^2$, $\Omega_{\mathrm{DM},0} h^2$, $100\,\theta_s$,
$\ln (10^{10} A_s)$, $n_s$, $\tau_{\mathrm{reio}}$, $w_\mathrm{f}$, $\beta$\}, \\
where $\Omega_{\mathrm{B},0} h^2$ and $\Omega_{\mathrm{DM},0} h^2$
are the baryon and cold dark matter densities relative to the
critical density respectively, $\theta_s$ is the ratio of the
sound horizon to the angular diameter distance at decoupling,
$A_s$ is the amplitude of the primordial scalar perturbation
spectrum, $n_s$ is the scalar spectral index,
$\tau_{\mathrm{reio}}$ is the optical depth to reionization,
$w_\mathrm{f}$ is the fluid equation of state parameter, and
$\beta$ is the Tsallis parameter. Furthermore, we have four
derived parameters which are the reionization redshift
($z_\mathrm{reio}$), the matter density parameter
($\Omega_{\mathrm{M},0}$), the Hubble constant ($H_0$), and the
root-mean-square mass fluctuations on scales of 8 $h^{-1}$ Mpc
($\sigma_8$). According to preliminary numerical works, we
consider the prior range [$0.98$, $1.008$] for the Tsallis
parameter, and also we set no prior range on the fluid equation of state parameter $w_\mathrm{f}$.

In this analysis we use six likelihoods: The Planck likelihood
with Planck 2018 data (containing high-$l$ TT,TE,EE, low-$l$ EE,
low-$l$ TT, and lensing) \citep{p18}, the Planck-SZ likelihood for
the Sunyaev-Zeldovich effect measured by Planck \citep{sz1,sz2},
the CFHTLenS likelihood with the weak lensing data
\citep{lens1,lens2}, the Pantheon likelihood with the supernovae data
\citep{pan},  the BAO likelihood with the baryon acoustic
oscillations (BAO) data \citep{bao1,bao2}, and the BAORSD likelihood for
BAO and redshift-space distortions (RSD) measurements
\citep{rsd1,rsd2}.

\begin{figure}
 \centering
 \includegraphics[width=8.5cm]{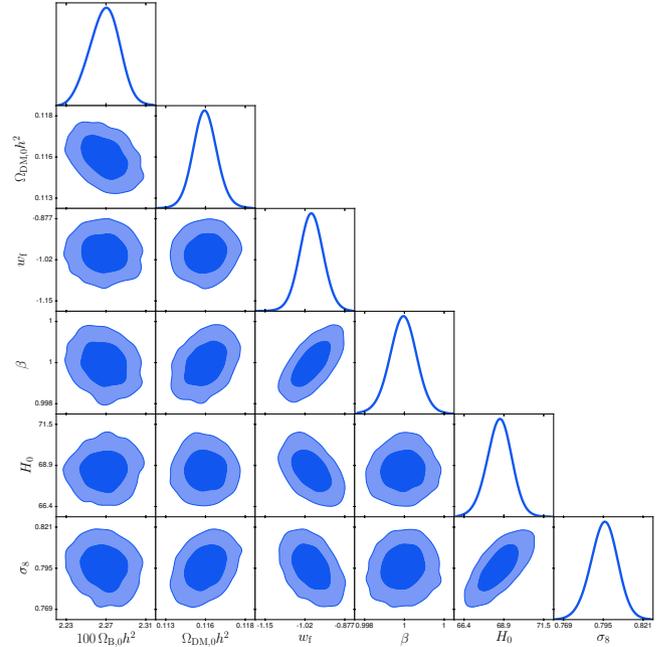}
 \caption{\small The one-dimensional posterior distribution and two-dimensional posterior contours with 68\% and 95\% confidence limits from the "Planck + Planck-SZ + CFHTLenS + Pantheon + BAO + BAORSD" data set, for the selected cosmological parameters of TMG model.}
 \label{f4}
 \end{figure}

Constraints on the cosmological parameters, considering the
combined "Planck + Planck-SZ + CFHTLenS + Pantheon + BAO + BAORSD" data
set, are displayed in Table \ref{t1}.
    \begin{table}
        \centering
        \caption{Best fit values and 68\% and 95\% confidence
            limits for cosmological parameters from "Planck + Planck-SZ +
            CFHTLenS + Pantheon + BAO + BAORSD" data set for $\Lambda$CDM and TMG
            model.}
        \scalebox{.64}{
            \begin{tabular}{|c|c|c|c|c|}
                \hline
                & \multicolumn{2}{|c|}{} & \multicolumn{2}{|c|}{} \\
                & \multicolumn{2}{|c|}{$\Lambda$CDM} & \multicolumn{2}{|c|}{TMG model} \\
                \cline{2-5}
                & & & & \\
                {parameter} & best fit & 68\% \& 95\% limits & best fit & 68\% \& 95\% limits \\ \hline
                & & & & \\
                $100\,\Omega_{\mathrm{B},0} h^2$ & $2.261$ & $2.263^{+0.012+0.026}_{-0.013-0.025}$ & $2.272$ & $2.268^{+0.014+0.027}_{-0.015-0.028}$ \\
                & & & & \\
                $\Omega_{\mathrm{DM},0} h^2$ & $0.1163$ & $0.1164^{+0.00078+0.0015}_{-0.00079-0.0015}$ & $0.1166$ & $0.1160^{+0.00097+0.0016}_{-0.00077-0.0017}$ \\
                & & & & \\
                $100\,\theta_s$ & $1.042$ & $1.042^{+0.00029+0.00055}_{-0.00026-0.00053}$ & $1.042$ & $1.042^{+0.00026+0.00054}_{-0.00028-0.00052}$ \\
                & & & & \\
                $\ln (10^{10} A_s)$ & $3.034$ & $3.024^{+0.010+0.023}_{-0.014-0.021}$ & $3.026$ & $3.028^{+0.011+0.024}_{-0.014-0.025}$ \\
                & & & & \\
                $n_s$ & $0.9712$ & $0.9719^{+0.0036+0.0072}_{-0.0039-0.0074}$ & $0.9706$ & $0.9720^{+0.0038+0.0074}_{-0.0040-0.0076}$ \\
                & & & & \\
                $\tau_\mathrm{reio}$ & $0.05358$ & $0.04963^{+0.0041+0.010}_{-0.0074-0.0096}$ & $0.05102$ & $0.05158^{+0.0059+0.011}_{-0.0072-0.012}$ \\
                & & & & \\
                $w_{\mathrm{f}}$ & --- & --- & $-0.9677$ & $-0.9944^{+0.041+0.089}_{-0.046-0.084}$ \\
                & & & & \\
                $\beta$ & --- & --- & $0.9999$ & $0.9997^{+0.00047+0.00098}_{-0.00048-0.00090}$ \\
                & & & & \\
                $z_\mathrm{reio}$ & $7.502$ & $7.084^{+0.50+1.0}_{-0.69-1.0}$ & $7.231$ & $7.275^{+0.59+1.2}_{-0.72-1.2}$ \\
                & & & & \\
                $\Omega_{\mathrm{M},0}$ & $0.2871$ & $0.2876^{+0.0043+0.0086}_{-0.0044-0.0086}$ & $0.2994$ & $0.2945^{+0.0076+0.016}_{-0.0074-0.015}$ \\
                & & & & \\
                $H_0\;[\mathrm{km\,s^{-1}\,Mpc^{-1}}]$ & $69.56$ & $69.54^{+0.37+0.73}_{-0.36-0.71}$ & $68.20$ & $68.62^{+0.79+1.7}_{-0.89-1.8}$ \\
                & & & & \\
                $\sigma_8$ & $0.8079$ & $0.8044^{+0.0045+0.0096}_{-0.0051-0.0091}$ & $0.7927$ & $0.7956^{+0.0097+0.019}_{-0.0092-0.019}$ \\
                & & & & \\
                \hline
            \end{tabular}
        }
        \label{t1}
    \end{table}
The triangle plot for selected cosmological parameters of TMG
model is illustrated in Fig. \ref{f4}.
According to our results, TMG model predicts a lower value for the
structure growth parameter $\sigma_8$ with respect to $\Lambda$CDM
model. So it seems that TMG model is more compatible with
local measurements of structure growth, with a slight
alleviation of $\sigma_8$ tension. Also the derived best fit value
of $w_\mathrm{f}$ indicates that the fluid "f" has a negative
pressure, and so we can conclude that it is a dark energy fluid.
Interestingly enough, the best fit and the mean value of
the fluid equation of state parameter demonstrate the
quintessential character of dark energy, which is also effective
in reducing the growth of structure. However, the phantom nature
of $w_\mathrm{f}$ is also permitted within the $1\sigma$
confidence level, where $-1.040<w_\mathrm{f}<-0.9532$. Moreover,
according to the correlation between $\sigma_8$ and $H_0$, one can
conclude that TMG model is not capable of providing a full
reconciliation between low redshift probes and CMB measurements.

Furthermore, in order to understand which model provides a better
fit to observational data, one can use the Akaike information
criterion (AIC) defined as \citep{aic1,aic2}
    \begin{equation}
    \mathrm{AIC}=-2\ln{\mathcal{L}_{\mathrm{max}}}+2K ,
    \end{equation}
in which $\mathcal{L}_{\mathrm{max}}$ is the maximum likelihood
function, and $K$ is the number of free parameters. According to
numerical analysis we have $\mathrm{AIC_{(\Lambda CDM)}}=3847.12$
and $\mathrm{AIC_{(TMG)}}=3850.84$, which results in
$\Delta\mathrm{AIC}=3.72$. So it can be concluded that the
$\Lambda$CDM model is more supported by observational data, while
the TMG model can not be ruled out.
\section{Conclusions} \label{sec5}
According to the analogy between thermodynamics and gravity, one
can rewrite the Friedmann equation of FLRW universe in the form of
the first law of thermodynamics at apparent horizon and vice
versa. On the other hand, the Boltzmann-Gibbs (BG) theory is not
convincing in gravitational systems which leads us to choose
non-additive entropy to study such systems. In the present paper,
we have considered the non-additive Tsallis entropy expression at
apparent horizon of FLRW universe, to disclose its novel features
on cosmological parameters. Accordingly we have derived the
gravitational field equations when the entropy associated with the
horizon is in the form of Tsallis entropy given in equation
(\ref{eq2}). In the cosmological background the obtained TMG leads
to modified Friedmann equations describing the evolution of the
universe. Employing numerical analysis, we constrain the
cosmological parameters of this model with observational data,
based on Planck CMB, weak lensing, supernovae, BAO, and RSD data.
The obtained results indicate that the preferred value of
$\sigma_8$ in TMG model is lower than the one in $\Lambda$CDM, and
so TMG model with a quintessential dark energy can slightly reduce
the well known tension between the Planck CMB and local
observations of $\sigma_8$.
\section*{Acknowledgments}
We are grateful to anonymous referee for valuable
comments which helped us improve the paper significantly. We also
thank Shiraz University Research Council.
\section*{Data availability}
No new data were generated or analysed in support of this research.

\bibliographystyle{mnras}
\bibliography{1}

\bsp
\label{lastpage}
\end{document}